\documentstyle[prb,aps,12pt,preprint]{revtex}

\begin{document}
\draft
\title{Magnetic field effects on the density of states of orthorhombic
superconductors}
\author{E. Schachinger and J.P. Carbotte}
\address{Department of Physics and Astronomy, McMaster University\\
  Hamilton, Ont. L8S 4M1\\
   Canada
}
\date{\today}
\maketitle
\begin{abstract}
The quasiparticle density of states in a two-dimensional $d$-wave
superconductor depends on the orientation of the in-plane external magnetic
field {\bf H}. This is because, in the region of the gap nodes, the
Doppler shift due to the circulating supercurrents around a vortex
core depend on the direction of {\bf H}. For a tetragonal system
the induced pattern is four-fold symmetric and, at zero energy, the
density of states exhibits minima along the node directions. But
YBa$_2$C$_3$O$_{6.95}$ is orthorhombic because of the chains and the
pattern becomes two-fold symmetric with the position of the minima
occuring when {\bf H} is oriented along the Fermi velocity at
a node on the Fermi surface. The effect of impurity
scattering in the Born and unitary limit are discussed.
\end{abstract}
\pacs{74.20.Fg 74.25.Ha 74.72.-h}
\newpage
\section{Introduction}

It has become increasingly clear that including the circulating
supercurrents outside the vortex core in a $d$-wave superconductor
can lead to a qualitative understanding - in the low magnetic field
limit, near $H_{c1}$ - of the effect of a magnetic field {\bf H} on
various properties of a superconductor in the vortex state.
The $\sqrt{H}$ dependence of the specific heat
predicted by Volovik\cite{Volo1,Volo2} and Kopnin and Volovik\cite{%
Kopn1,Kopn2} was verified experimentally.\cite{Moler1,Moler2,Junod}
A detailed analysis of the data for {\bf H} oriented perpendicular
to the CuO$_2$ planes including impurity scattering was given by
K\"ubert and Hirschfeld\cite{Kub1} (see also Barash {\it et al.}%
\cite{Barash}).
Transport properties were considered by K\"ubert and Hirschfeld.%
\cite{Kub2} Subsequently the analysis has been extended by Hirschfeld%
\cite{Hirsch1} to include the field {\bf H} in the CuO$_2$ plane and
he demonstrated a four-fold symmetry in the electronic thermal
conductivity. Very recently results have been obtained by Vekhter
{\it et al.}\cite{Vek1,Vek2} for the effect of an in-plane magnetic field on the
quasiparticle density of states (QDOS) in a tetragonal two dimensional
$d$-wave superconductor. The present work extends this previous study 
to the orthorhombic $d$-wave case and also includes in the calculations 
the effect of impurity scattering in both, Born's and resonant scattering 
limits.

While most high $T_c$ oxides are tetragonal, YBa$_2$Cu$_3$O$_{6.95}$ (YBCO),
which is perhaps the most extensively studied of all the oxides, is
orthorhombic because its crystal structure includes CuO chains oriented
along the $b$-axis, as well as two CuO$_2$ planes per unit cell. Optical
measurements reveal that the carrier density in the chains in optimally
doped YBCO is of the same order of magnitude as it is in the planes and
the ratio of the associated $b$- to $a$-direction plasma frequency is
of the order of 2.\cite{Tanner} Similar large in-plane anisotropies
are observed in the infrared optical conductivity,\cite{Basov} d.c.\
resistivity,\cite{Gagnon1,Fried} zero temperature penetration depth,%
\cite{Zhang,Bonn} and thermal conductivity.\cite{Yu,Gagnon2}

Various approaches have been taken to treat band anisotropy including
a single tight binding band with different nearest neighbor hopping
in $a$- and $b$-direction,\cite{Odonov1,Odonov2} two band\cite{Atkin1,%
Atkin2,Odonov3,Odonov4} as well as three band\cite{Odonov5,Atkin3} models
which include the single CuO chain layer as well as two CuO$_2$ plane
layers per unit cell with transverse hopping between them. All these
models give the same qualitative features and describe reasonably well
the $a$-$b$ band anisotropy of YBCO. Additionally, in an orthorhombic structure the
superconducting gap will be a mixture\cite{Odonov6} of dominant
$d_{x^2-y^2}$ symmetry as well as subdominant $s_0$ and $s_{x^2+y^2}$
symmetric functions because they all belong to the same irreducible
representation of the $C_{2v}$ crystal point group.

The simplest approach which contains the physics of the more complex
calculations is a single infinite band with distinct effective masses
along the $a$ $(m_a)$ and $b$ $(m_b)$ directions\cite{Kim,Wu,Schur,Ju}
with a $(d+s)$-symmetric gap function on the cylindrical Fermi surface.%
\cite{Maki,Modre} For simplicity we will adopt throughout this paper
this approach and calculate the effect of an in-plane magnetic field {\bf H} on
the QDOS assuming a distribution of non overlapping vortex cores and account only
for the circulating supercurrents around the vortices. We will also
treat the effect of impurities in both, Born's and
resonant scattering limits. Impurities were not considered in the
previous calculations in tetragonal symmetry by Vekhter {\it et al.}\cite{Vek1}
which were confined to the pure limit. Thus, our results with impurities
in the tetragonal case are the first such results.

\section{Formalism}

We begin with a single free electron band in two dimensions with an
ellipsoidal Fermi surface and the single particle energy
$\varepsilon_{\bf k}$ in the normal state given by the dispersion
relation $(\hbar = 1)$
\begin{equation}
 \varepsilon_{\bf k} = {k_x^2\over 2m_x}+{k_y^2\over 2m_y}-
  \varepsilon_F  \label{e1}
\end{equation}
where ${\bf k} = (k_x,k_y)$ is the momentum vector and $\varepsilon_F$
the Fermi energy. In Eq.~(\ref{e1}) $m_x$ and $m_y$ are effective
masses with $m_x > m_y$, {\it i.e.} chains are along the $k_y$-axis
which increases the electronic transport in that direction. A parameter
\begin{equation}
  \alpha = {m_x-m_y\over m_x+m_y}  \label{e2}
\end{equation}
is introduced and this is the single parameter which characterizes the
band structure anisotropy in our simple model. Consistent with the
above assumption for the energy band dispersion relation is a mixed
$(d+s)$-symmetric gap on the Fermi surface (an ellipse) of the form
\begin{equation}
  \Delta_{\bf k} = \Delta_0\left (\cos 2\varphi +s\right ),
  \label{e3}
\end{equation}
with $\Delta_0$ the gap amplitude. In Eq.~(\ref{e3}) $\varphi$ is a
polar angle measured from the $k_x$-axis in the two dimensional copper
oxide Brillouin zone giving the angle of {\bf k} (momentum on the
Fermi surface) as shown in Fig.~\ref{f1}. The first term in Eq.~(\ref{e3})
gives the dominant $d$-wave part of the gap which, on its own, would have
nodes on the main diagonal at $\varphi = \pm\pi/4$. The number $s$ is
a measure of the subdominant $s$-wave part. Both parts are allowed to
exist in an orthorhombic system such as YBCO from group theoretical
considerations. The $s$-component moves the nodes from
$\varphi = \pm\pi/4$ to $\varphi = \tan^{-1}\sqrt{(1+s)/(1-s)}$ off the
main diagonal of the Brillouin zone. For positive values of $s$ the
critical $\varphi$ defining the node is larger than $\pi/4$ while the opposite
holds for negative values of $s$.

In order to work with Eqs.~(\ref{e1}) and (\ref{e3}) it is convenient
to transform to new coordinates ${\bf p} = (p_x,p_y)$ in which the
Fermi surface is a circle (see Fig.~\ref{f1}).\cite{Kim} The required
transformation is given by
\begin{equation}
 p_i = k_i\sqrt{m_x+m_y\over 2m_i},\qquad i = x,y
 \label{e4}
\end{equation}
which leads to (Fig.~\ref{f1})
\begin{equation}
  \tan\phi = \sqrt{m_y\over m_x}\tan\varphi,
  \label{e5}
\end{equation}
where $\phi$ is now the angle of the momentum vector {\bf p} in the
transformed frame. Clearly, the electron dispersion relation (\ref{e1})
reduces to
\begin{equation}
 \varepsilon_F = {p^2_{F,x}+p^2_{F,y}\over 2\bar{m}} =
   {p^2_F\over 2\bar{m}},
 \label{e6}
\end{equation}
with $\bar{m} = (m_x+m_y)/2$ the average band mass. Applying the same
transformation to the gap (\ref{e3}) gives the order parameter in the
{\bf p}-frame\cite{Wu,Schur,Ju}
\begin{equation}
  \Delta_{\b p} = \Delta_0\left({\alpha+\cos 2\phi\over 1+\alpha
    \cos 2\phi}+s\right) \equiv \Delta_0 f(\phi).
  \label{e7}
\end{equation}

We show in Fig.~\ref{f2} the gap $\Delta_{\bf p}$ in the {\bf p}-frame for
several values of $\alpha$ (the band anisotropy) and $s$ (the gap
anisotropy). Frame (a) is for $\alpha = 0$ and $s=0$ and
is the well known pure $d$-wave case for which
plus and minus lobes have equal absolute magnitude and the zeros are
on the diagonal. In frame (b) $\alpha = 0.4$ and $s=0$, {\it i.e.} the
gap in the {\bf k}-frame (laboratory frame) is of pure $d$-wave
symmetry but the band structure orthorhombicity leads to a transformed
gap with larger positive than negative lobes and the nodes are shifted
upwards from $\pm\pi/4$ and similarly for the other symmetry related
angles. This trend is even more pronounced in frame (c) where
$\alpha = s = 0.4$. In this case $\alpha$ and $s$ add constructively to
further increase the positive lobes as compared to the negative ones,
and the nodes are shifted to yet larger angles. Finally, frame (d)
applies to $\alpha = 0.4$ and $s=-0.4$. For this set of parameters the
nodes in the {\bf p}-frame remain on the diagonal but at the same time
the negative lobes are dominant over the positive ones. We conclude that
a negative value for $s$ can in some sense counteract some of the effects
of a positive $\alpha$, but there is no complete cancelation between
these two parameters which play quite distinctive roles. The nodes can be
made to remain unshifted but the size of the lobes is changed over their
$\alpha = s = 0$ values.

In the laboratory frame ({\bf k}, before transformation) the
thermodynamic Green's function takes on the form
\begin{equation}
  G({\bf k},\omega_n;{\bf r}) = -{\left(i\omega_n-{\bf v}_F({\bf k})
  {\bf q}_s\right)\tau_0 +\varepsilon_{\bf k}\tau_3+\Delta_{\bf k}\tau_1
  \over \left(\omega_n+i{\bf v}_F({\bf k}){\bf q}_s\right)^2+
  \varepsilon^2_{\bf k} +\Delta^2_{\bf k}},
 \label{e8}
\end{equation}
with $\varepsilon_{\bf k}$ and $\Delta_{\bf k}$ given by Eqs.~(\ref{e1})
and (\ref{e3}) respectively. In Eq.~(\ref{e8}) the $\tau$'s are Pauli
$2\times 2$ matrices and $i\omega_n$ is the $n$'th Matsubara frequency,
$\omega_n = (2n+1)\pi T, n = 0,\pm1,\pm2,\ldots$, and $T$ is the
temperature. The Doppler shift caused by the presence of circulating
supercurrents outside the vortex core (the region which we assume to
dominate the physics) is given by ${\bf v}_F({\bf k}){\bf q}_s$, where
${\bf v}_F({\bf k})$ is the electron Fermi velocity on the ellipsoidal
Fermi surface in {\bf k}-space (see Fig.~\ref{f1}) and which is
given by
\begin{equation}
  {\bf v}_F({\bf k}) = \nabla_{\bf k}\varepsilon_{\bf k} =
   {k_{F,x}\over m_x}{\bf\hat{x}}+{k_{F,y}\over m_y}{\bf\hat{y}},
  \label{e9}
\end{equation}
with ${\bf \hat{x}}$ and ${\bf \hat{y}}$ unity vectors in the direction of
the $k_x$- and $k_y$-axis. Applying the transformation (\ref{e4}) to get
${\bf v}_F({\bf k})$ in the {\bf p}-frame leads to
\begin{equation}
  {\bf v}_F({\bf p}) = {1\over\sqrt{\bar{m}}}\left(
   {p_{F,x}\over\sqrt{m_x}}{\bf\hat{x}} +
   {p_{F,y}\over\sqrt{m_y}}{\bf\hat{y}}\right).
   \label{e10}
\end{equation}

The momentum of the superfluid currents outside the vortex core (placed
at the position ${\bf r} = 0$ in the laboratory frame) is assumed to be
inversely proportional to the distance $\vert{\bf r}\vert = r$ from the
core. For a magnetic field {\bf H} in the CuO$_2$ plane making an
angle $\gamma$ with the $k_x$-axis the supercurrents are in a plane
perpendicular to {\bf H} and therefore to the $k_x$-$k_y$ plane.
Assuming circular supercurrents, an approximation discussed in the paper
by Vekhter {\it et al.},\cite{Vek1} and denoting the vortex winding
angle by $\beta$ one obtains for the $x$ and $y$ component of the
superfluid momentum ${\bf q}_s$
\begin{eqnarray}
   q_{s,x} &=& -\vert {\bf q}_s\vert \sin\beta\sin\gamma,\nonumber\\
   q_{s,y} &=& -\vert {\bf q}_s\vert \sin\beta\cos\gamma,
   \label{e11}
\end{eqnarray}
so that
\begin{equation}
 {\bf v}_F({\bf p}){\bf q}_s =
  {p_F\over m_x}\vert{\bf q}_s\vert\sin\beta\sqrt{1+\alpha}\left (
  \sqrt{1+\alpha\over 1-\alpha}\sin\phi\cos\gamma-\cos\phi\sin\gamma
  \right).
 \label{e12}
\end{equation}
If one measures the distance $r$ in units of the intervortex distance 
$R = a^{-1}\sqrt{\phi_0\over\pi H}$, where $\phi_0$ is the flux quantum
and $a$ a constant of the order of unity, we obtain
\begin{equation}
 {\bf v}_F({\bf p}){\bf q}_s =
  {E_H\over\rho}\sin\beta\sqrt{1+\alpha}\left (
  \sqrt{1+\alpha\over 1-\alpha}\sin\phi\cos\gamma-\cos\phi\sin\gamma
  \right),
 \label{e13}
\end{equation}
with the normalized distance $\rho = r/R$ and the magnetic energy
\begin{equation}
  E_H = {a\over 2}\bar{v}_F\sqrt{\pi H\over\phi_0} = \nu\Delta_0,
  \label{e14}
\end{equation}
with $\bar{v}_F$ a suitably defined effective Fermi velocity\cite{Vek1}
which can also be made to contain the additional band factor
$\sqrt{1+\alpha}$ of Eq.~(\ref{e13}) and $\nu$ is a parameter which
measures the magnetic energy $E_H$ in units of the gap. Consequently,
we can write
\begin{equation}
 {\bf v}_F({\bf p}){\bf q}_s = {E_H\over\rho}h(\phi,\gamma)\sin\beta,
 \label{e15}
\end{equation}
with
\begin{equation}
  h(\phi,\gamma) = 
     \sqrt{1+\alpha\over 1-\alpha}\sin\phi\cos\gamma-\cos\phi\sin\gamma.
  \label{e16}
\end{equation}

The QDOS at zero frequency obtained from the Green's function (\ref{e8})
which includes the circulating supercurrents about the vortex core
and which must also be averaged over the vortex unit cell is given by
\begin{equation}
 N(0,\gamma) = \int\limits^{2\pi}_0\!{d\phi\over 2\pi}\,
   \int\limits^{2\pi}_0\!{d\beta\over\pi}\,
   \int\limits^R_0\!{r dr\over R^2}\,\Re{\rm e}\left\{
   {\vert {\bf v}_F({\bf p}){\bf q}_s\vert\over
    \sqrt{({\bf v}_F({\bf p}){\bf q}_s)^2-\Delta^2_0f^2(\phi)}}
   \right\}.
 \label{e17}
\end{equation}
By the square root is meant that branch of the complex function which corresponds
to a positive imaginary part.
Eq.~(\ref{e17}) can be reduced to an one dimensional
integral form
\begin{equation}
  N(0,\gamma) = {1\over 2\pi}\int\limits^{2\pi}_0\!d\phi\,
   {\rm min}\left( 1,{\nu^2\vert h(\phi,\gamma)\vert^2\over
    f^2(\phi)} \right).
  \label{e18}
\end{equation}
One good approximation to this integral is obtained by noting that the
integrand in Eq.~(\ref{e18}) is dominated by the nodal region and, thus,
we get approximately
\begin{equation}
  N(0,\gamma) \simeq 2{\nu\over\pi}\sum\limits_{\rm nodes}
  \left\vert {h(\phi_n,\gamma)\over f'(\phi_n)}\right\vert,
  \label{e19}
\end{equation}
where the $\phi_n$ are the nodes of the transformed gap (\ref{e7}) and
$f'(\phi_n)$ is the derivative of this gap function at its nodes. These
can be found at
\begin{equation}
   \tan\phi_n = \pm\sqrt{{(1+\alpha)\over(1-\alpha)}
     {(1+s)\over(1-s)}},
   \label{e20}
\end{equation}
while in the laboratory frame, as we have noted before, the gap has its
nodes at $\tan\varphi_n = \pm\sqrt{(1+s)/(1-s)}$. Furthermore, the
derivative $f'(\phi_n)$ at the gap nodes can be worked out to be
\begin{equation}
  f'(\phi_n) = 2(1+s)\sqrt{(1-s^2)(1-\alpha^2)},
  \label{e21}
\end{equation}
and hence an approximation for $N(0,\gamma)$ follows from Eqs.~(\ref{e18}),
(\ref{e21}) and the definition of $h(\phi,\gamma)$ given in Eq.~%
(\ref{e15}) as
\begin{equation}
 N(0,\gamma) = {\nu\over\pi}\sum\limits_{\rm nodes}
  {\left\vert\sqrt{1+\alpha\over 1-\alpha}\sin\phi_n\cos\gamma-
  \cos\phi_n\sin\gamma\right\vert\over
  (1+s)\sqrt{(1-s^2)(1-\alpha^2)}},
  \label{e22}
\end{equation}
with distinct contributions from only two of the four nodes.

For one set of two nodes we get
\begin{eqnarray}
   \cos\phi_n &=& \sqrt{{1\over 2}{(1-\alpha)(1-s)\over 1+\alpha s}},
    \nonumber\\
   \sin\phi_n &=& \sqrt{{1\over 2}{(1+\alpha)(1+s)\over 1+\alpha s}},
  \label{e23}
\end{eqnarray}
while for the other ones $\phi_n\to\pi-\phi_n$. We therefore arrive at the
approximate analytic expression
\begin{eqnarray}
  N(0,\gamma) &=& {\sqrt{2}\nu\over\pi}{1\over\sqrt{(1-s^2)(1-\alpha^2)}}
   {1\over 1+s}\sqrt{1+\alpha\over 1-\alpha}\nonumber\\
  &&\times\left\{\left\vert(1+\alpha)\sqrt{1+s\over1+\alpha s}\cos\gamma-
    (1-\alpha)\sqrt{1-s\over 1+\alpha s}\sin\gamma\right\vert\right .
     \nonumber\\
  &&+\left.\left\vert(1+\alpha)\sqrt{1+s\over 1+\alpha s}\cos\gamma+
      (1-\alpha)\sqrt{1-s\over 1+\alpha s}\sin\gamma\right\vert\right\}
      \nonumber\\
  &=& {2\sqrt{2}\nu\over\pi}{1\over\sqrt{(1-s^2)(1-\alpha^2)}}
      {1\over 1+s}\sqrt{1+\alpha\over 1-\alpha}\nonumber\\
  && \times{\rm max}\left(
      \left\vert(1+\alpha)\sqrt{1+s\over 1+\alpha s}\cos\gamma\right\vert,
      \left\vert(1-\alpha)\sqrt{1-s\over 1+\alpha s}\sin\gamma\right\vert
      \right),
  \label{e24}
\end{eqnarray}
where we have restored factors previously absorbed in the magnetic energy
$E_H$ that appeared explicitely in Eq.~(\ref{e13}). In the tetragonal,
pure $d$-wave case $(\alpha = s = 0)$ expression (\ref{e24}) reduces
properly to the known result\cite{Vek1}
\begin{equation}
  N(0,\gamma) = {2\sqrt{2}\nu\over\pi}{\rm max}\left(\vert\sin\gamma\vert,
   \vert\cos\gamma\right\vert),
  \label{e25}
\end{equation}
in which case Eq.~(\ref{e25}) is known to be an excellent
approximation to the one dimensional integral of Eq.~(\ref{e18}).

When band anisotropy is included through a finite value of $\alpha$ or
a subdominant $s$-wave part, characterized by the parameter $s$,
the analytic expression (\ref{e24}) is not quite as good 
an approximation as (\ref{e25}) was but it is
still acceptable as can be seen in Fig.~\ref{f3}. In this figure we
compare with our exact numerical results for $N(0,\gamma)$ based on the
one dimensional integral of Eq.~(\ref{e18}). Here only band anisotropy
is included with an effective mass anisotropy parameter $\alpha = 0.2$.
The dashed line is the exact result (\ref{e18}), the solid line is based
on Eq.~(\ref{e24}). The agreement is good. Note the two-fold symmetry
in the pattern obtained, although, since $s=0$, the gap in the laboratory
frame has pure $d$-wave symmetry with nodes on the main diagonal.
 It is the band anisotropy that breaks the four-fold symmetry of the
tetragonal case. However, it is
very important to realize that because of $\alpha\ne 0$,
{\it i.e.} band orthorhombicity, the
sharp kinks in the curve for $N(0,\gamma)$ vs.\ $\gamma$ no longer fall
at the position of the nodes. This would be the case in a tetragonal
system but is not so for the ellipsoidal Fermi surface of Fig.~\ref{f1}.
We see from Eq.~(\ref{e24}) that the kink occurs at a critical angle
$\gamma_c$ given by (note that $s=0$):
\begin{equation}
  \tan\gamma_c = {1+\alpha\over 1-\alpha},
  \label{e26}
\end{equation}
which is $\pi/4$ only for the case $\alpha = 0$ (no anisotropy in the
electronic band structure). This is an important result because it is
often believed that the sharp structure in $N(0,\gamma)$ as a function
of $\gamma$ - the magnetic field orientation - gives the position of
the nodes. This is not true when band anisotropy exists.

The geometry of the situation is shown in Fig.~\ref{f4} which depicts
the laboratory frame while the result (\ref{e24}) was obtained in the
transformed, {\bf p}-frame where the Fermi surface is a circle. In this
figure ${\bf k}_n$ gives the position of a gap node on the ellipsoidal
Fermi surface $\varepsilon_{\bf k} = 0$, in Eq.~(\ref{e1}). The Fermi
velocity at that point, ${\bf v}_F({\bf k}_n)$, points in the direction
$\tan^{-1}\left({1+\alpha\over 1-\alpha}\tan\varphi\right)$. If the
magnetic field {\bf H} is placed in that direction the circulating
supercurrents will be in a plane perpendicular to the two dimensional
Brillouin zone and oriented along the tangent to the Fermi surface at
${\bf k} = {\bf k}_n$ as shown in the figure by the thick shaded line.
 Thus, there will be no
Doppler shift. In the laboratory frame $\tan\varphi = \sqrt{(1+s)/(1-s)}$
from Eq.~(\ref{e3}) and we get immediately the result
\begin{equation}
  \tan\gamma_c = {1+\alpha\over 1-\alpha}\sqrt{1+s\over 1-s}
  \label{e27}
\end{equation}
for a finite subdominant $s$-wave part to the gap. This same result
follows from Eq.~(\ref{e24}) when the expression in the absolute
value switches sign. For finite values of $s$, the angle $\gamma_c$
coincides with the position of the nodes in the $(d+s)$-symmetric
gap in the laboratory frame only if $\alpha = 0$. We can also see
from (\ref{e27}) that a negative value of the subdominant $s$-wave part
({\it i.e.} a negative value of $s$) partially compensates for the effect
of $\alpha$ but does not cancel it out completely.

In Fig.~\ref{f4New} we show further results for the zero frequency
QDOS $N(0,\gamma)$ as a function of magnetic field orientation $\gamma$
for increasingly anisotropic bands. The parameter $\alpha$ is increased
from 0.2 in the solid curve to 0.4 (dashed curve), 0.6 (dotted
curve), and 0.8 (dashed-dotted curve). It is clear that the amplitude
of the predicted oscillations in the QDOS increases considerably with increasing
effective mass anisotropy $(\alpha)$ and that consequently this anisotropy
will be easier to detect as the orthorhombicity of the band structure
increases, {\it i.e.} as $\alpha$ increases. There is nearly a factor of
6 between the value at minimum and at maximum in the dashed-dotted curve.
The pattern of oscillation is also changed as $\alpha$ increases. The
intermediate secondary maximum, around $\gamma = 1.5\,{\rm rad}$ in the
solid curve, is not present any longer in the dashed-dotted curve.
Instead there is a minimum.

A fit to data of the form (\ref{e24}) could in principle yield information
on the two most important parameters $\alpha$ and $s$ but the position
of the sharp structure in the QDOS by itself does not give directly the
position of the nodes or the ratio of the subdominant $s$-wave admixture
to the dominant $d$-wave component. It is the direction of {\bf H}
giving no Doppler shift which determines the position of the minimum
in $N(0,\gamma)$. Before leaving Eq.~(\ref{e24}) we note that
$N(0,\gamma)$ is proportional to $\nu = E_H/\Delta_0$ and this is the same
dependence on the magnetic field as has been found in the tetragonal
case. There is no explicit change in this quantity due to anisotropy so that
 the scaling
variable for the QDOS and for the specific heat stays unchanged. This
holds for the pure case but will be modified when disorder is considered.
In that case scaling breaks down\cite{Moler2} and specific details
play a role.

The finite frequency QDOS in a pure orthorhombic superconductor with a
$(d+s)$-symmetric gap is also of interest. It follows from the formula
\begin{equation}
 N(\omega,\gamma) = \int\limits^{2\pi}_0\!{d\beta\over\pi}\,
   \int\limits^1_0\!d\rho\,\rho
   \int\limits^{2\pi}_0\!{d\phi\over 2\pi}\,
   \Re{\rm e}\left\{
   {\tilde{\omega}(\omega)-{\bf v}_F({\bf p}){\bf q}_s\over
    \sqrt{(\tilde{\omega}(\omega)-{\bf v}_F({\bf p}){\bf q}_s)^2
    -\tilde{\Delta}^2_{d+s}(\omega,\phi)}}
   \right\},
 \label{e28}
\end{equation}
which has been written in a form which remains valid even when impurity
scattering is included. Here $\rho = r/R$. For the pure case
 $\tilde{\omega}(\omega) =
\omega$ and $\tilde{\Delta}_{d+s}(\omega,\phi) = \Delta_0f(\phi)$.
Results of the numerical evaluation of Eq.~(\ref{e28}) (an approximate
analytic formula similar to Eq.~(\ref{e24}) can also be derived but
is not given here as it has the same form as already presented in the
work of Vekhter {\it et al.}\cite{Vek1,Vek2}) gives the results shown
in Fig.~\ref{f5} for four directions of the magnetic field.
Here we present numerical results for $N(\omega,\gamma)$
as a function of $\omega/\Delta_0$ for $\alpha = 0.2$ and $\nu=0.3$.
For the magnetic field in the direction of $\gamma = 0^\circ$ (antinodal
direction in pure $d$-wave) the solid curve applies and an $\omega^2$
behavior is obtained at small $\omega$, just as in the tetragonal case.\cite{%
Vek1,Vek2} But the linear dependence expected in the nodal direction
in the tetragonal case now occurs in some other direction where the
Doppler shift at the nodal position is zero, namely at $\gamma_c$
as defined by Eq.~(\ref{e27}) which is where the superfluid momentum
${\bf q}_s$ is perpendicular to the Fermi velocity. For $\alpha=0.2$,
$\gamma_c\simeq 56^\circ$ the dashed-dotted curve indeed displays
linear dependence.

\section{Impurity Effects}

We now include in the calculations elastic impurity scattering. In this
case the full self consistent Green's function (\ref{e8}) enters the
impurity term and we need to solve two coupled equations, one for the
renormalized Matsubara frequency $\tilde{\omega}(\omega)$ and the other
one for the gap $\tilde{\Delta}_{d+s}(\omega,\phi)$. On the real
frequency axis they are:
\begin{mathletters}
\label{e29}
\begin{eqnarray}
 \tilde{\omega}(\rho,\beta,\omega) &=& \omega+i\pi t^+
   \Omega(\rho,\beta,\omega), \label{e29a}\\
 \tilde{\Delta}_{d+s}(\rho,\beta,\omega,\phi) &=&
  \Delta_{d+s}(\phi)+i\pi t^+ D(\rho,\beta,\omega), \label{e29b}
\end{eqnarray}
\end{mathletters}
with $\Delta_{d+s}(\phi) = \Delta_{\bf p}$ of Eq.~(\ref{e7}),
the pure system gap in the transformed frame. Here $\rho$ and $\beta$
are associated with the vortex and $\phi$ is the polar angle defining
the direction of {\bf p}. The use of Eqs.~(\ref{e29}) confines our
analysis to low temperatures where the temperature dependence of the
gap function can be neglected. Moreover, these equations are only
valid in Born's approximation, or weak impurity scattering limit. A
slightly more complicated set of equations needs to be solved in the
unitary limit (strong impurity scattering), namely
\begin{mathletters}
\label{e30}
\begin{eqnarray}
 \tilde{\omega}(\rho,\beta,\omega) &=& \omega+i\pi \Gamma^+
   {\Omega(\rho,\beta,\omega)\over\Omega^2(\rho,\beta,\omega)+
    D^2(\rho,\beta,\omega)}, \label{e30a}\\
 \tilde{\Delta}_{d+s}(\rho,\beta,\omega,\phi) &=&
  \Delta_{d+s}(\phi)+i\pi \Gamma^+ {D(\rho,\beta,\omega)\over
  \Omega^2(\rho,\beta,\omega)+D^2(\rho,\beta,\omega)}. \label{e30b}
\end{eqnarray}
\end{mathletters}
In the tetragonal case $(\alpha = 0)$, also considered here, the
symmetry is described by the C$_{4v}$ point group and the $(d+is)$-symmetry
of the order parameter is the only possible mixed symmetry allowed
under the symmetry operations of this group. In this case Eqs.~(\ref{e29b})
and (\ref{e30b}) are to be modified to give
\begin{equation}
 \tilde{\Delta}_s(\rho,\beta,\omega) =
  \Delta_s+i\pi t^+ D(\rho,\beta,\omega), \label{e29c}
\end{equation}
in Born's limit and
\begin{equation}
 \tilde{\Delta}_s(\rho,\beta,\omega) =
  \Delta_s+i\pi \Gamma^+ {D(\rho,\beta,\omega)\over
  \Omega^2(\rho,\beta,\omega)+D^2(\rho,\beta,\omega)}. \label{e30c}
\end{equation}
in the resonant scattering limit. $\Delta_s$ is the $s$-wave contribution
to the order parameter in the clean limit and is equal to zero in
pure $d$-wave symmetry. The renormalized gap function is then written as:
\begin{equation}
  \tilde{\Delta}_{d+is}(\rho,\beta,\omega,\phi) =
   \Delta_d(\phi)+i\tilde{\Delta}_s(\rho,\beta,\omega).
   \label{e30d}
\end{equation}
Finally, the functions $D(\rho,\beta,\omega)$ and $\Omega(\rho,\beta,\omega)$
are defined as
\begin{mathletters}
\label{e31}
\begin{eqnarray}
\Omega(\rho,\beta,\omega) &=& \left\{\begin{array}{ll}
  \left\langle
  {\tilde{\omega}(\rho,\beta,\omega)-{\bf v}_F({\bf p}){\bf q}_s\over
  \sqrt{\left(\tilde{\omega}(\rho,\beta,\omega)-{\bf v}_F({\bf p}){\bf q}_s
   \right)^2-\tilde{\Delta}^2_{d+s}(\rho,\beta,\omega,\phi)}}
  \right\rangle_\phi & \quad {\rm orthorhombic,}\\
  \left\langle
  {\tilde{\omega}(\rho,\beta,\omega)-{\bf v}_F({\bf p}){\bf q}_s\over
  \sqrt{\left(\tilde{\omega}(\rho,\beta,\omega)-{\bf v}_F({\bf p}){\bf q}_s
   \right)^2-\tilde{\Delta}^2_s(\rho,\beta,\omega)-\Delta^2_d(\phi)}}
  \right\rangle_\phi & \quad {\rm tetragonal,}
  \end{array}
    \right .  \label{e31a}\\
 D(\rho,\beta,\omega) &=& \left\{\begin{array}{ll}
  \left\langle
  {\tilde{\Delta}_{d+s}(\rho,\beta,\omega,\phi)\over
  \sqrt{\left(\tilde{\omega}(\rho,\beta,\omega)-{\bf v}_F({\bf p}){\bf q}_s
   \right)^2-\tilde{\Delta}^2_{d+s}(\rho,\beta,\omega,\phi)}}
  \right\rangle_\phi & \quad {\rm orthorhombic,}\\
  \left\langle
  {\tilde{\Delta}_s(\rho,\beta,\omega)\over
  \sqrt{\left(\tilde{\omega}(\rho,\beta,\omega)-{\bf v}_F({\bf p}){\bf q}_s
   \right)^2-\tilde{\Delta}^2_s(\rho,\beta,\omega)-\Delta^2_d(\phi)}}
  \right\rangle_\phi & \quad {\rm tetragonal,}
  \end{array}
  \right .  \label{e31b}
\end{eqnarray}
\end{mathletters}
with $\langle\cdots\rangle_\phi$ the Fermi surface average in the
{\bf p}-frame. Finally, Eq.~(\ref{e28}) still applies for the QDOS and
it is only at this later stage that an average over vortex variables
is carried out. Eqs.~(\ref{e29}) or (\ref{e30}) need to be solved
self consistently for different values of $\rho$, $\beta$, and
$\omega$. The two remaining parameters are $t^+$ or $\Gamma^+$ which are
related to the strength of impurity scattering in the normal state
where $\pi t^+ = \pi\Gamma^+ = 1/2\tau_{imp}$ and $\tau_{imp}$ is the
 impurity scattering
time in the normal state. It should also be noted at this point that the
impurity scattering potential
in the laboratory frame is now anisotropic. The role
of anisotropic impurity scattering in anisotropic superconductors has been
studied by Hara\'{n} and Nagi.\cite{Nagi} According to their results the
influence of the impurity concentration on superconductivity varies with
the anisotropy of the scattering potential, qualitative changes in, say
the temperature dependence of the penetration depth or the energy
dependence of the optical conductivity, cannot be expected.

A first result with impurities is given in Fig.~\ref{f6} for Born
scattering with $t^+ = E_H/16$ in a tetragonal system with a pure
$d$-wave gap. Here, $\Delta_0 = 24\,{\rm meV}$, $\nu = 0.1$, and what is
plotted is the zero frequency value of the QDOS, $N(0,\gamma)$, as
a function of the orientation of the in-plane magnetic field {\bf H}
which makes an angle $\gamma$ with the ${\bf\hat{x}}$-axis. The solid
line includes impurities and is compared with the dotted line for the
pure case. We note that the minima are somewhat filled and rounded
by the impurities as we might have expected and the maxima are slightly
higher and broader. Impurities increase the QDOS in the low energy
region. Close examination of the two curves also shows that they both
have four-fold symmetry. This is to be expected, but it is pointed out
that, even for a tetragonal system, the introduction of impurities
introduces a finite $s$-wave component to the gap, which is now of
$(d+is)$-symmetry, in Eqs.~(\ref{e29c})
or (\ref{e30c}) because it is the self consistent Green's function
that enters the definition of (\ref{e31}) and this leads to a nonzero
value of $D(\rho,\beta,\omega)$.
Our final result remains four-fold symmetric, {\it i.e.} it
retains the full tetragonal symmetry.

The effect of impurity scattering is much larger in the unitary limit.
Results are shown in Fig.~\ref{f7}. Again the underlying system has
tetragonal symmetry and a pure $d$-wave gap. The impurity scattering
parameter is $\Gamma^+ = E_H/16$ with a gap amplitude $\Delta_0 =%
24\,{\rm meV}$ and $\nu = E_H/\Delta_0 = 0.1$ in Eqs.~(\ref{e30}).
The solid line represents our results for the value of the QDOS
$N(0,\gamma)$ as a function of the magnetic field orientation $\gamma$
in the CuO$_2$ plane. Notice the scale on the vertical axis and that
the size of $N(0,\gamma)$ is roughly three times its value for the
Born limit (Fig.~\ref{f6}). This changes quite a lot when the
frequency is increased. The dashed curve applies to
$\omega/\Delta_0 = 0.05$, the dotted one to $\omega/\Delta_0 = 0.1$,
and the dashed-dotted one to $\omega/\Delta_0 = 0.15$. The percent anisotropy
in the QDOS is reduced with increasing $\omega$.

Results for the orthorhombic case are shown in Figs.~\ref{f8} and
\ref{f9}. In these figures $\nu = E_H/\Delta_0 = 0.1$, the band
anisotropy $\alpha = 0.41$, and the $s$-wave admixture to the gap
is characterized by $s = -0.25$ taken from a previous fit to experimental
data in YBCO as discussed by Wu {\it et al.}\cite{Wu} and
Sch\"urrer {\it et al.}\cite{Schur} Reasonable values for the parameters
in our admittedly simplified band structure model, namely $\alpha =%
(m_x-m_y)/(m_x+m_y)$ and the gap anisotropy, namely $s$, can be set
in comparison with penetration depth data on YBCO at optimum doping.
The measured ratio of the zero temperature penetration depth depends
on the anisotropic effective masses and sets the value of $\alpha = 0.41$
which follows from $\lambda_a/\lambda_b = 1600\,$\AA$/1030\,$\AA.\cite{%
Schur,Maki,Modre,Bonn1} Furthermore, in our model the low temperature
slopes of the penetration depth are approximated by
\begin{eqnarray}
{d\lambda_{xx}\over dT} &\simeq& {2\ln(2)\over\Delta_0}(1-\alpha-s),
    \nonumber\\
{d\lambda_{yy}\over dT} &\simeq& {2\ln(2)\over\Delta_0}(1+\alpha+s).
\end{eqnarray}
The experiments of Bonn {\it et al.}\cite{Bonn1} give $s = -0.25$ as an
estimate of the orthorhombicity in the gap. It is these values that we
have used to estimate the expected anisotropy in the QDOS for {\bf H}
in the CuO$_2$ plane. As we have already stressed in the introduction,
the band structure in YBCO is much more complex than the model treated
here but the simplicity of our model allows us to get insight into the
role of orthorhombicity and develop a qualitative picture of its
effect. There are no adjustable parameters left.

For the orthorhombic case the pattern in Fig.~\ref{f8} for $N(0,\gamma)$
vs. $\gamma$ is two-fold symmetric even in the pure case. As before,
Born scattering (solid curve with $t^+ = E_H/16$) simply smooths out
slightly the minima seen in the pure case curve (dotted line). On the
other hand, unitary scattering, solid gray curve for $\Gamma^+ =%
E_H/16$, pushes the curve up by a factor of about two over the pure
limit case and the pattern is also significantly modified and smoother.
The small maxima around $\gamma = 1.5\,{\rm rad}$ and symmetry related
points in the pure curve (dotted) are completely washed out by unitary
impurity scattering. The amplitude of the variations in value of $N(0,\gamma)$
is also considerably reduced.

The curves described so far apply to untwinned single crystals. Three
more curves are shown in Fig.~\ref{f8}. The dashed-dotted curve
applies to the case of twinned samples and was obtained from the clean
limit (dotted curve) by averaging it with a similar curve displaced by
$90^\circ$. This corresponds to a simple average of equal numbers of
twins with ${\bf\hat{x}}$- and ${\bf\hat{y}}$-axis interchanged. It is
clear that this average greatly reduces the predicted anisotropy and
experiments aimed at discovering this anisotropy are best made on
untwinned samples. This is confirmed in preliminary data from
Junod's group\cite{Junod2} who indeed find no significant in-plane
anisotropy in the specific heat on twinned single crystals. At first
sight the noise in their experiment is below, but of the order of the
anisotropy predicted in our calculations. There are many reasons why
the actual anisotropy might in fact be even less than the amount
shown in our pure twinned curve (dashed-dotted). To mention only one:
YBCO is more three dimensional than many of the high $T_c$
oxides and this implies one more integration in the definition of the
QDOS (\ref{e28}) and this can be expected to reduce the anisotropy
further. The two other curves in Fig.~\ref{f8} include impurities and 
show even less anisotropy than does the pure case. The dashed curve applies
to Born scattering with $t^+ = E_H/16$ and the dashed gray curve to
resonant scattering with $\Gamma^+ = E_H/16$.

In Fig.~\ref{f9} we present finite frequency results for $N(\omega,\gamma)$
vs. $\gamma$ (the magnetic field orientation in the CuO$_2$ plane) for
four finite frequency values, namely $\omega/\Delta_0 = 0.05$
(dashed-dotted curve), $\omega/\Delta_0 = 0.1$ (dashed curve),
$\omega/\Delta_0 = 0.15$ (dash-double dotted curve), and
$\omega/\Delta_0 = 0.2$ (dotted curve). The results are to be compared
with the solid curve which applies to the case $\omega = 0$ and has
already been shown in Fig.~\ref{f7}. It is clear that the two
pronounced minima in the solid curve become shallower with increasing
$\omega$ and are already quite small for the dotted curve with
$\omega/\Delta_0 = 0.2$. This complicated pattern should be reflected
in the accompanying pattern for the temperature dependence of the
specific heat.

\begin{figure}
\caption{The frequency dependence $(\omega)$ in units of the gap amplitude
$\Delta_0$ of the QDOS, $N(\omega,\gamma)$, for various values of the
magnetic field orientation $\gamma$ and impurity content. The solid and
dashed lines are for the pure case. The dashed and dashed-dotted ones are
for $t^+ = E_H/16$ and the solid gray and dashed gray ones for
$\Gamma^+ = E_H/16$. In each pair of curves the first is for $\gamma =%
0^\circ$ while the second is in the direction where two node regions
have no Doppler shift due to the magnetic field. In (a) which applies
to the tetragonal case this is for $\gamma = 45^\circ$ which is the
nodal direction but in (b)  which applies to the orthorhombic case
the critical direction is for $\gamma = 61.3^\circ$. This is different
from the nodal direction in the laboratory frame.}
\label{f10}
\end{figure}
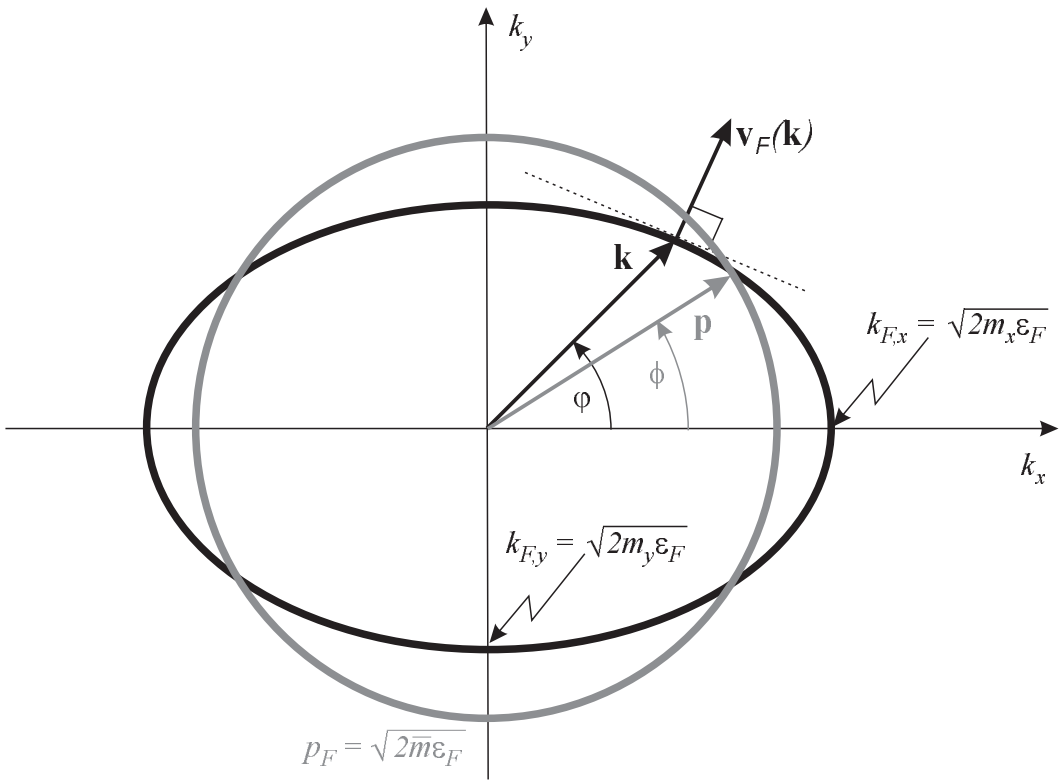
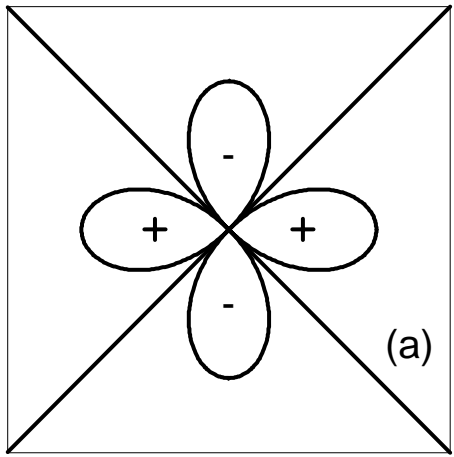
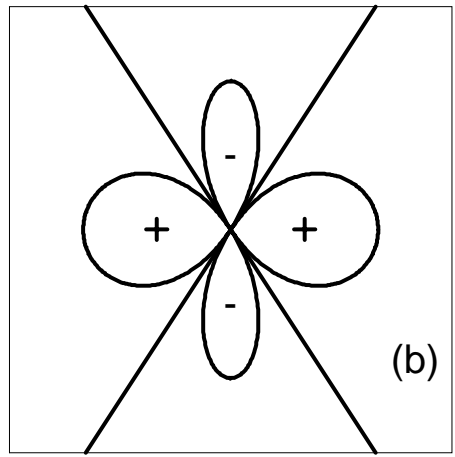
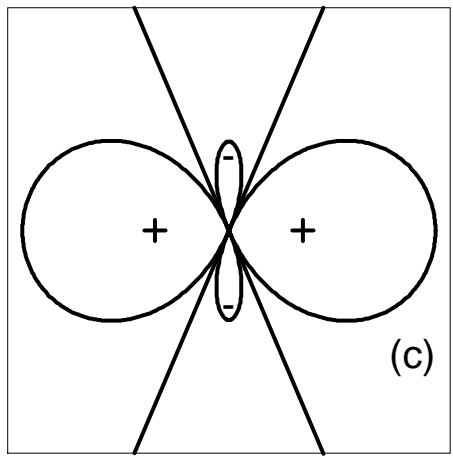
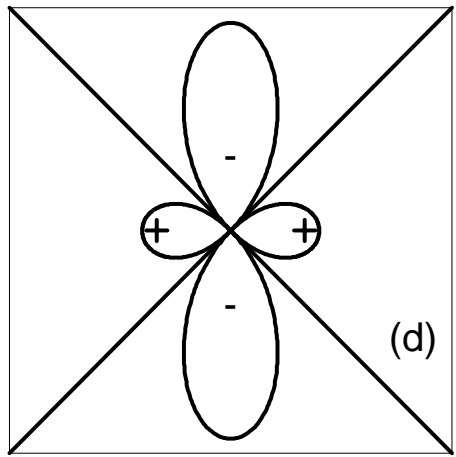
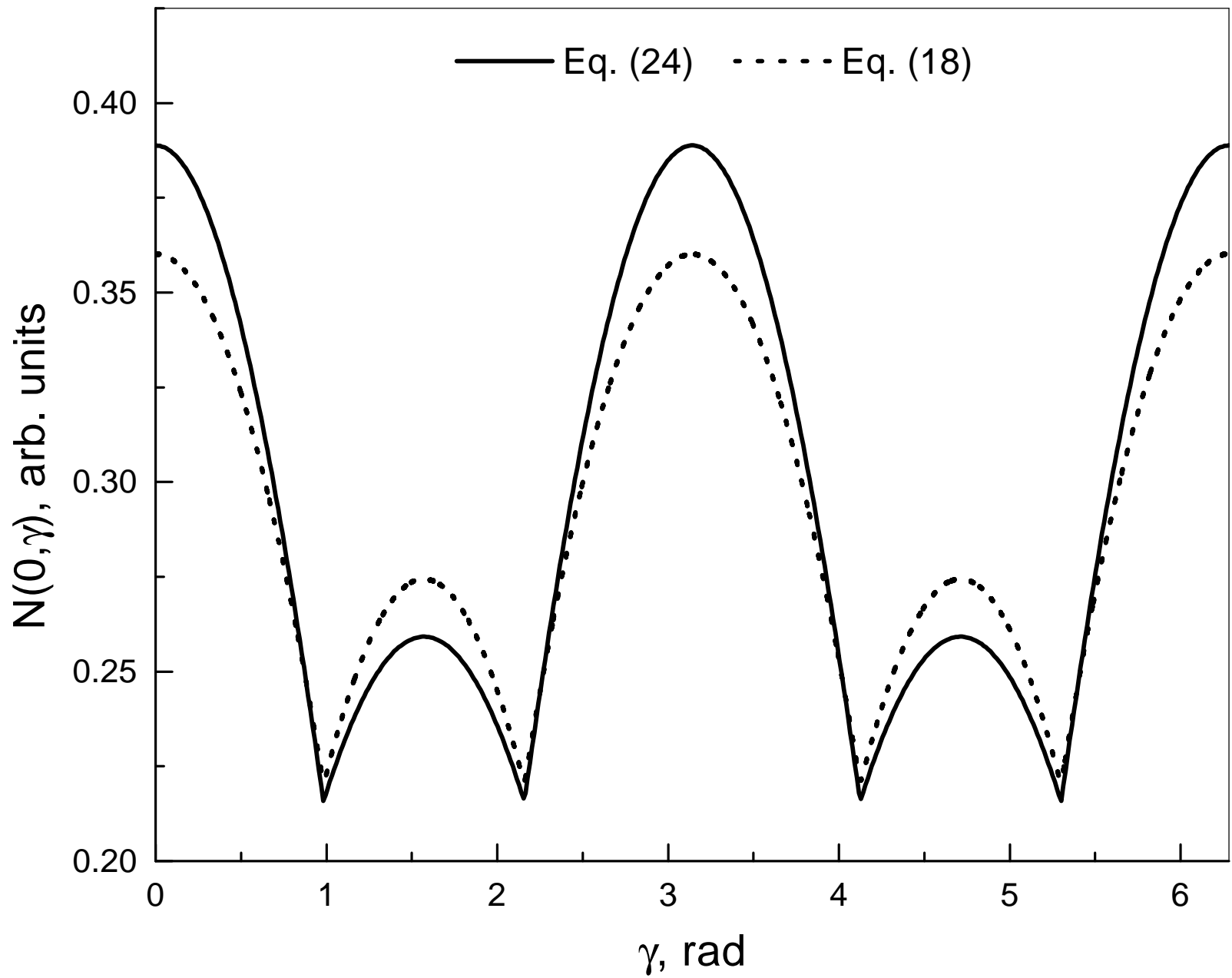
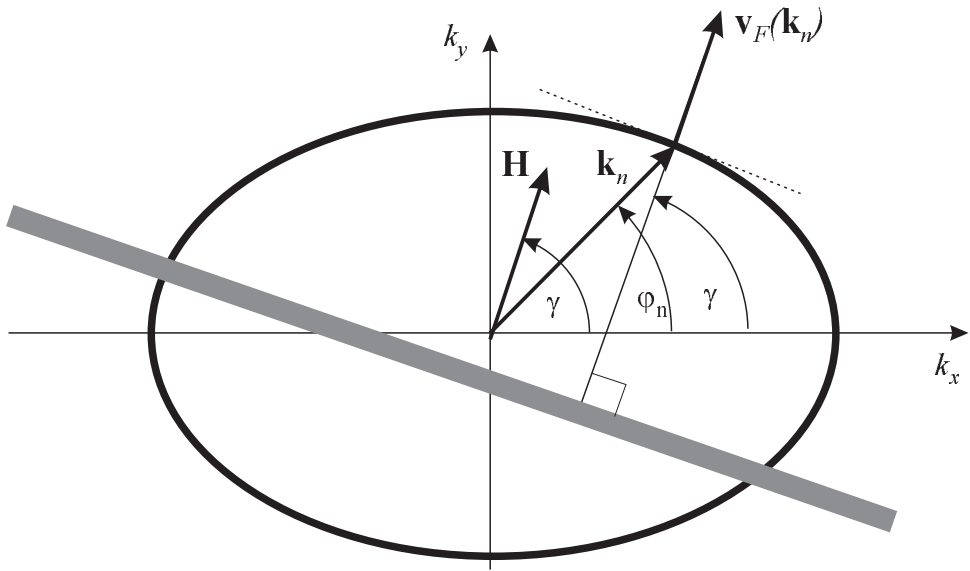
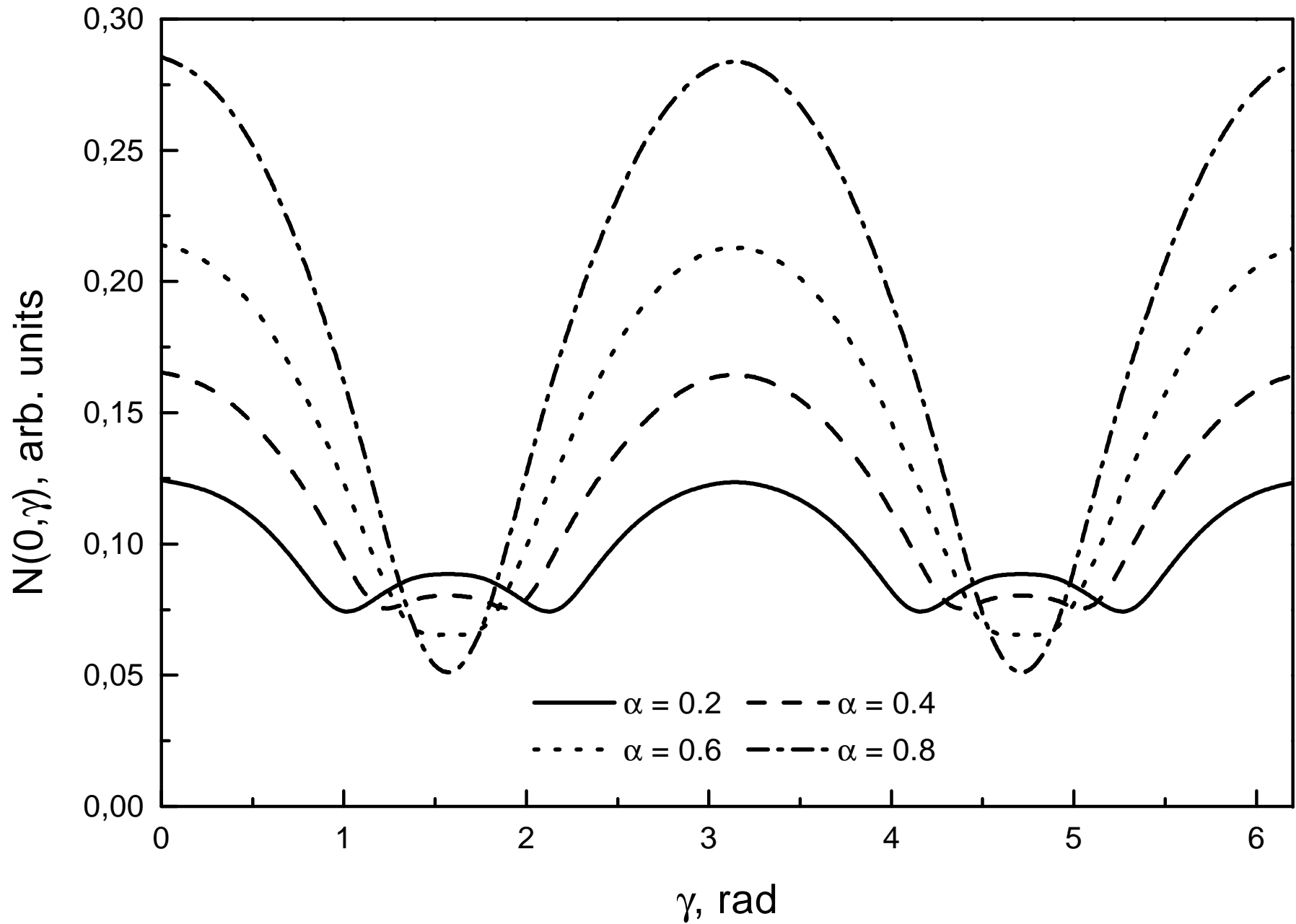
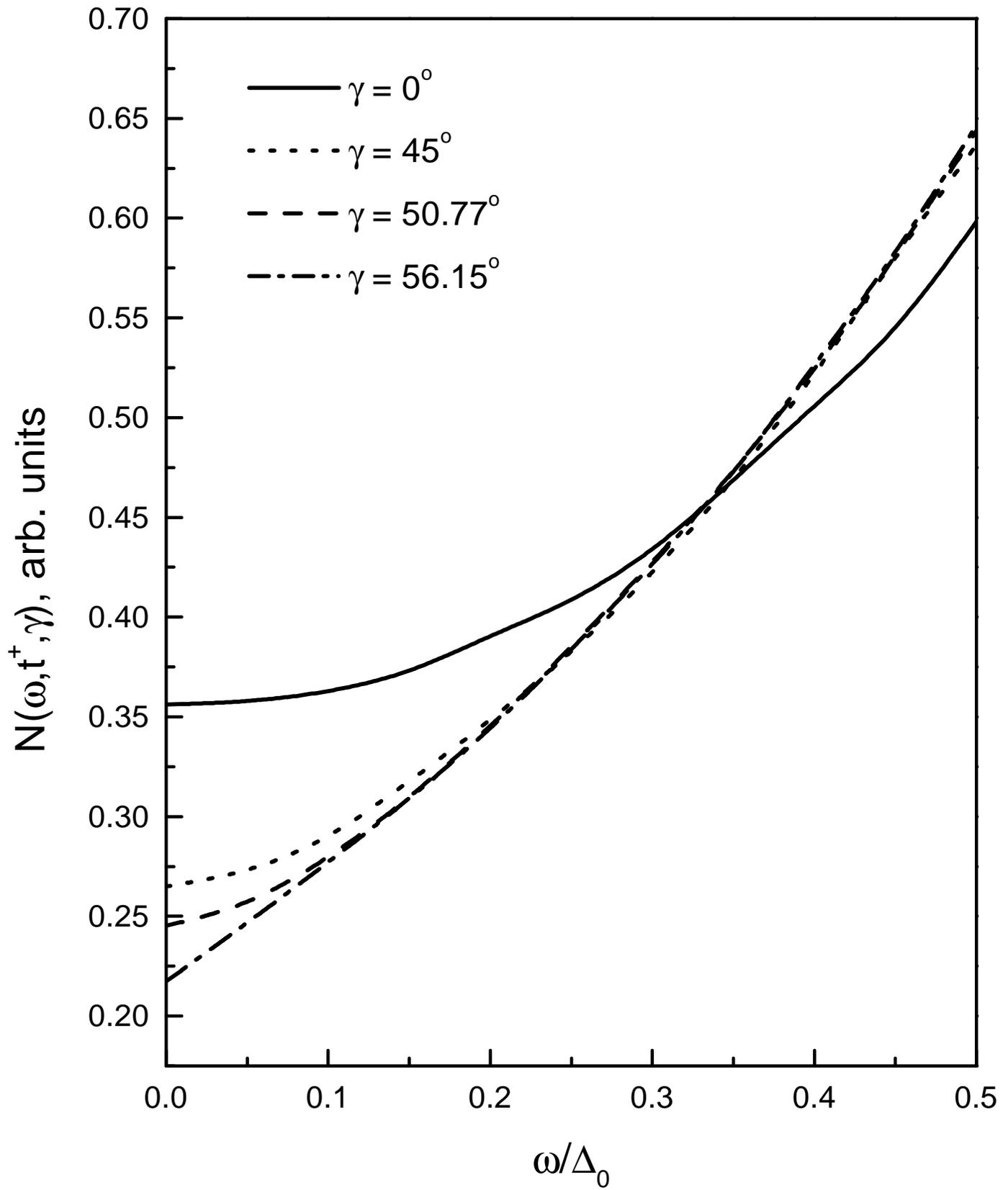
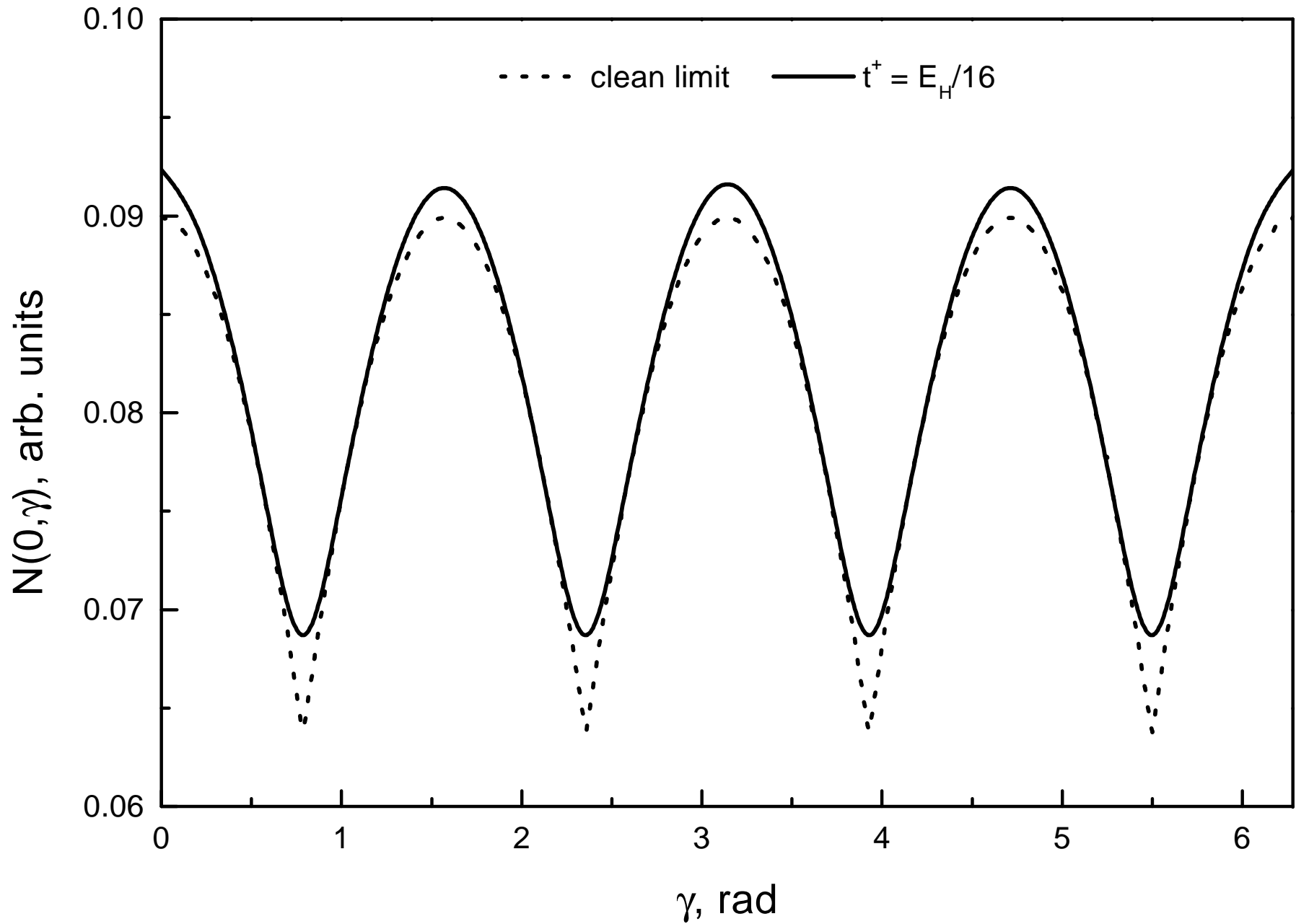
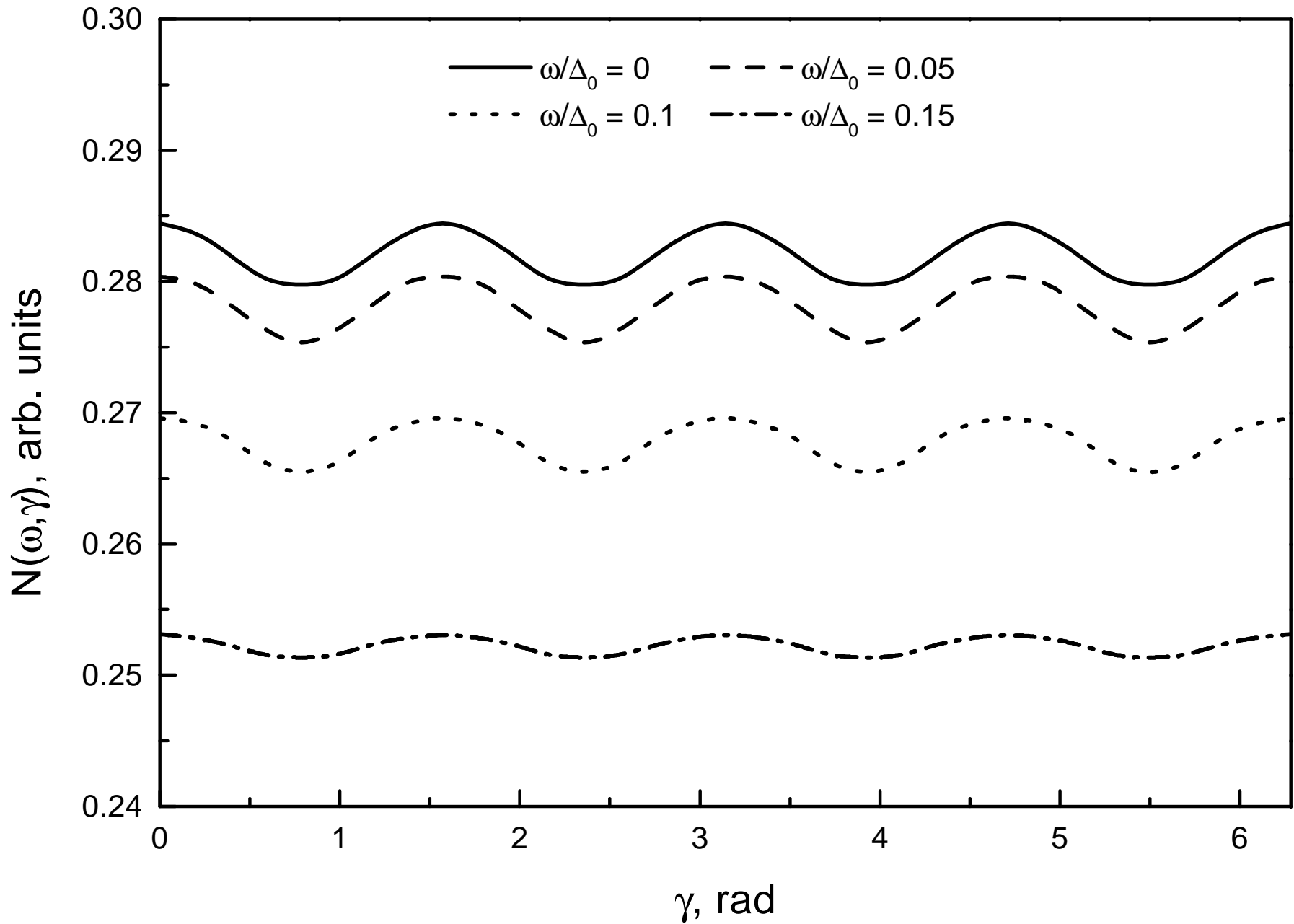
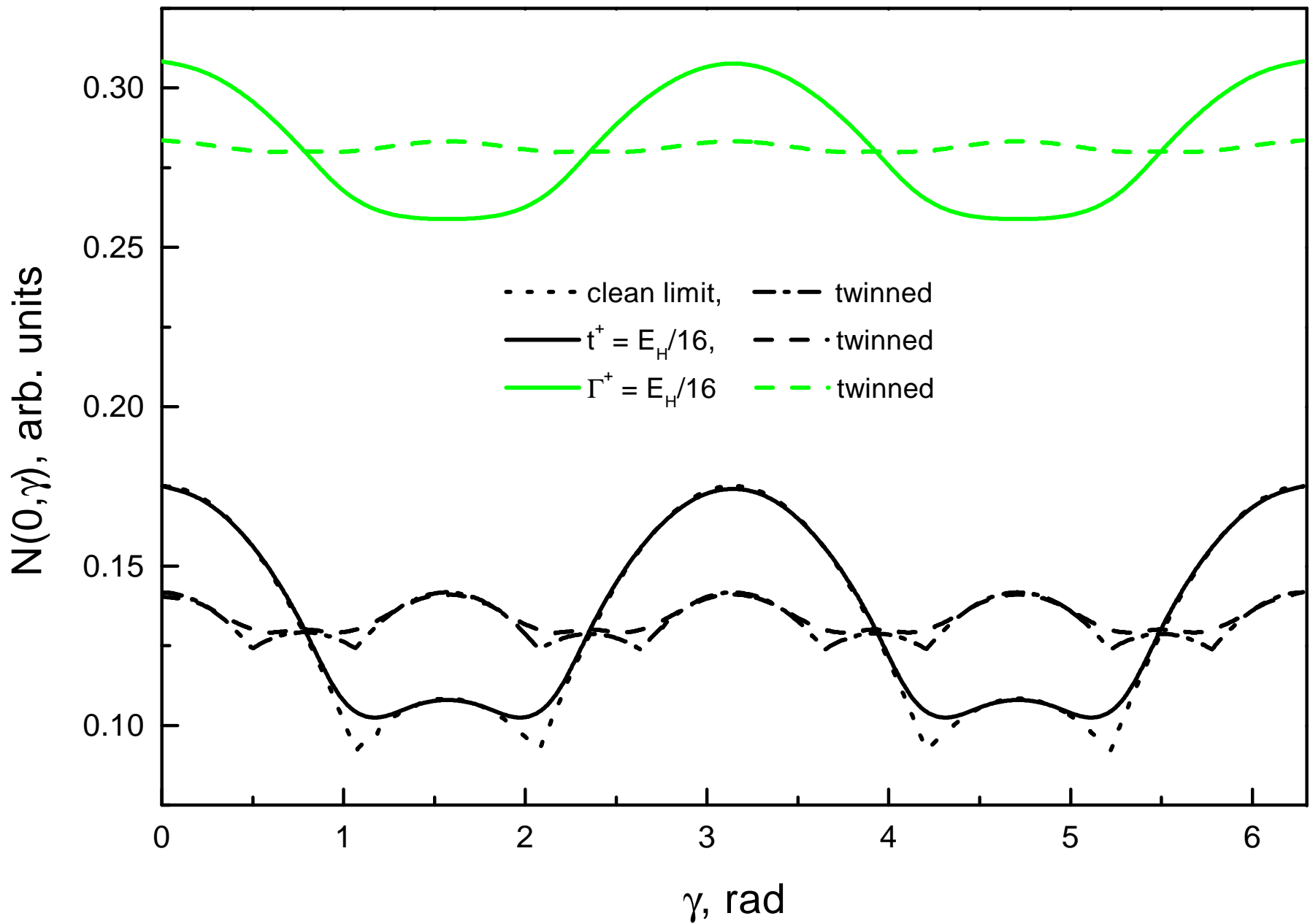
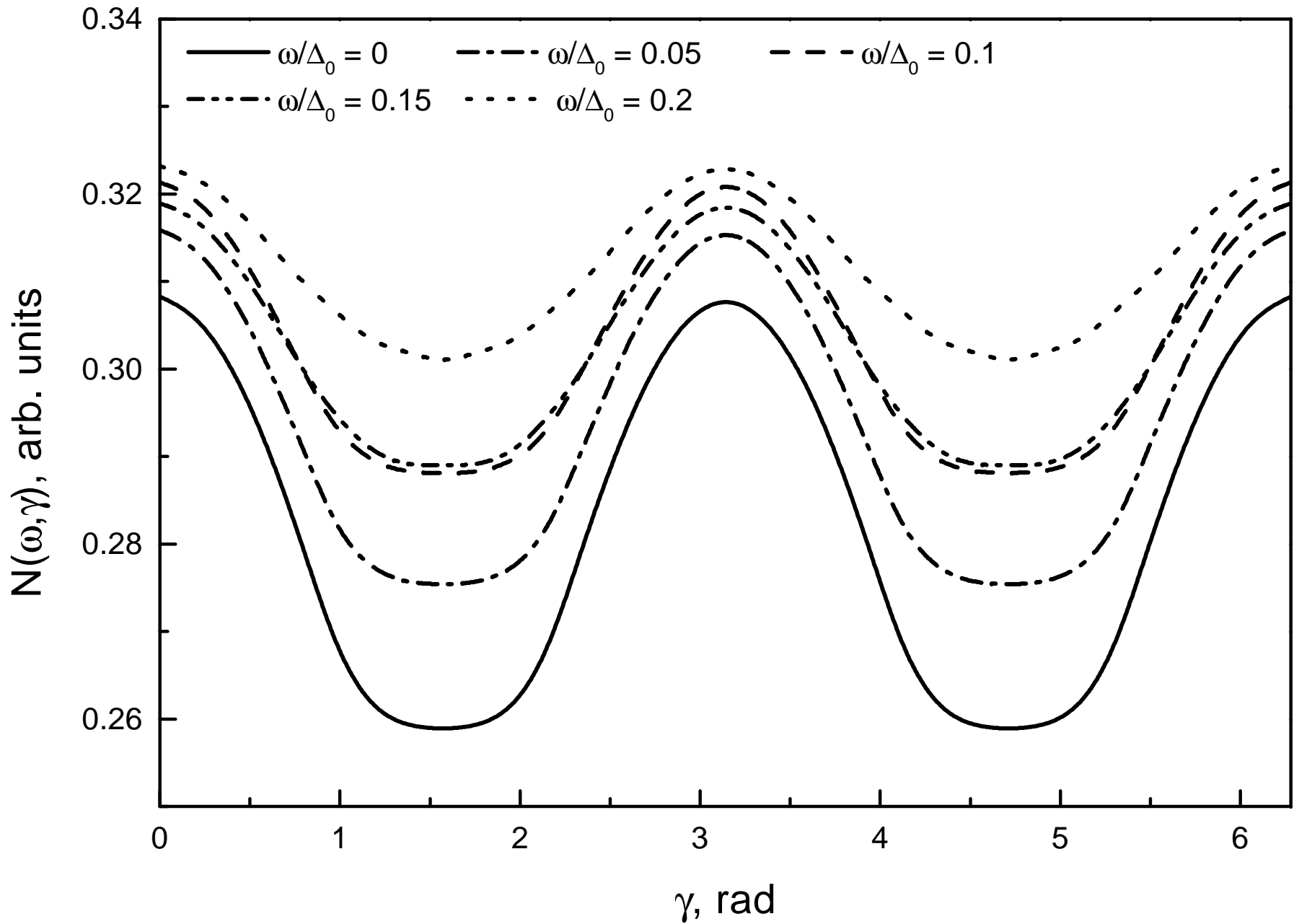
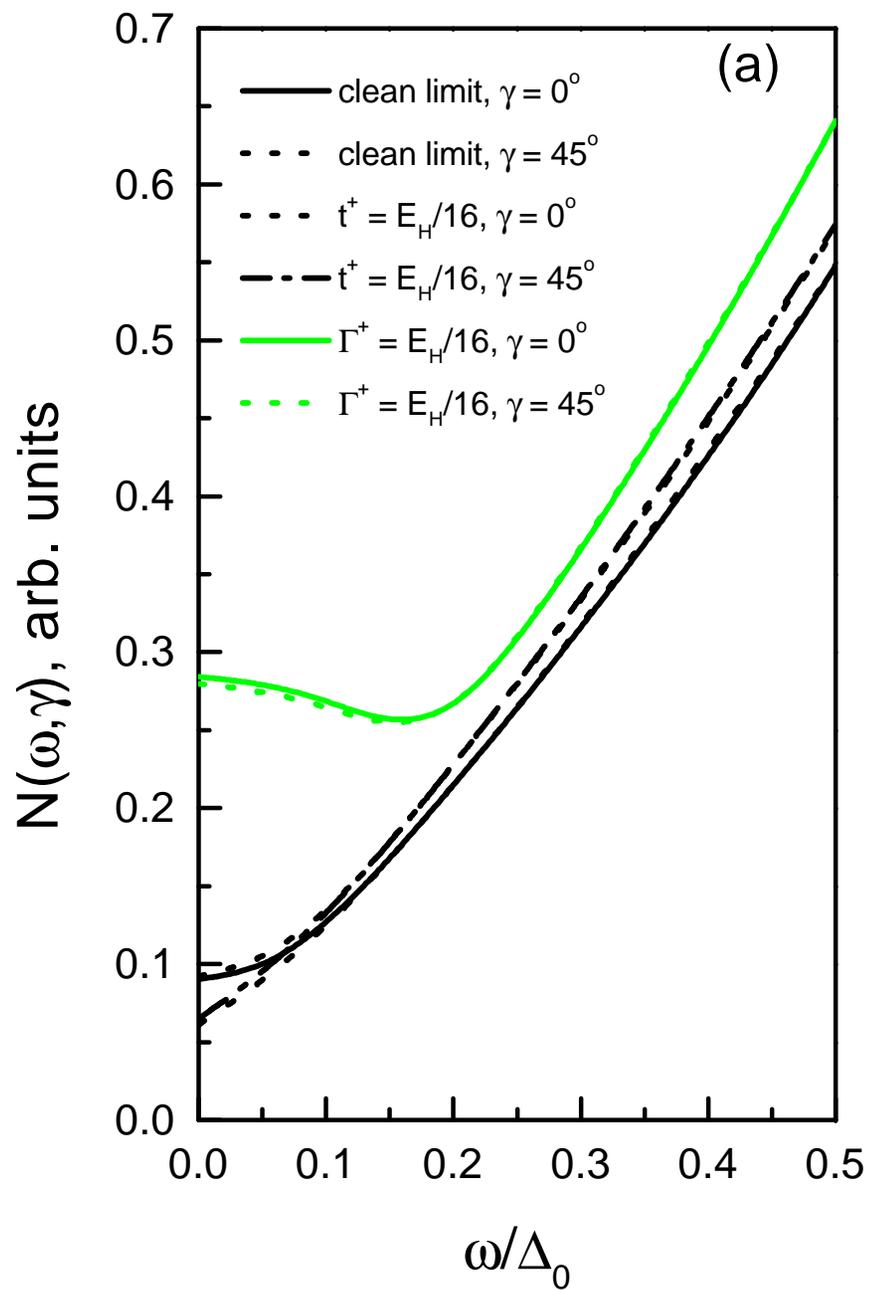
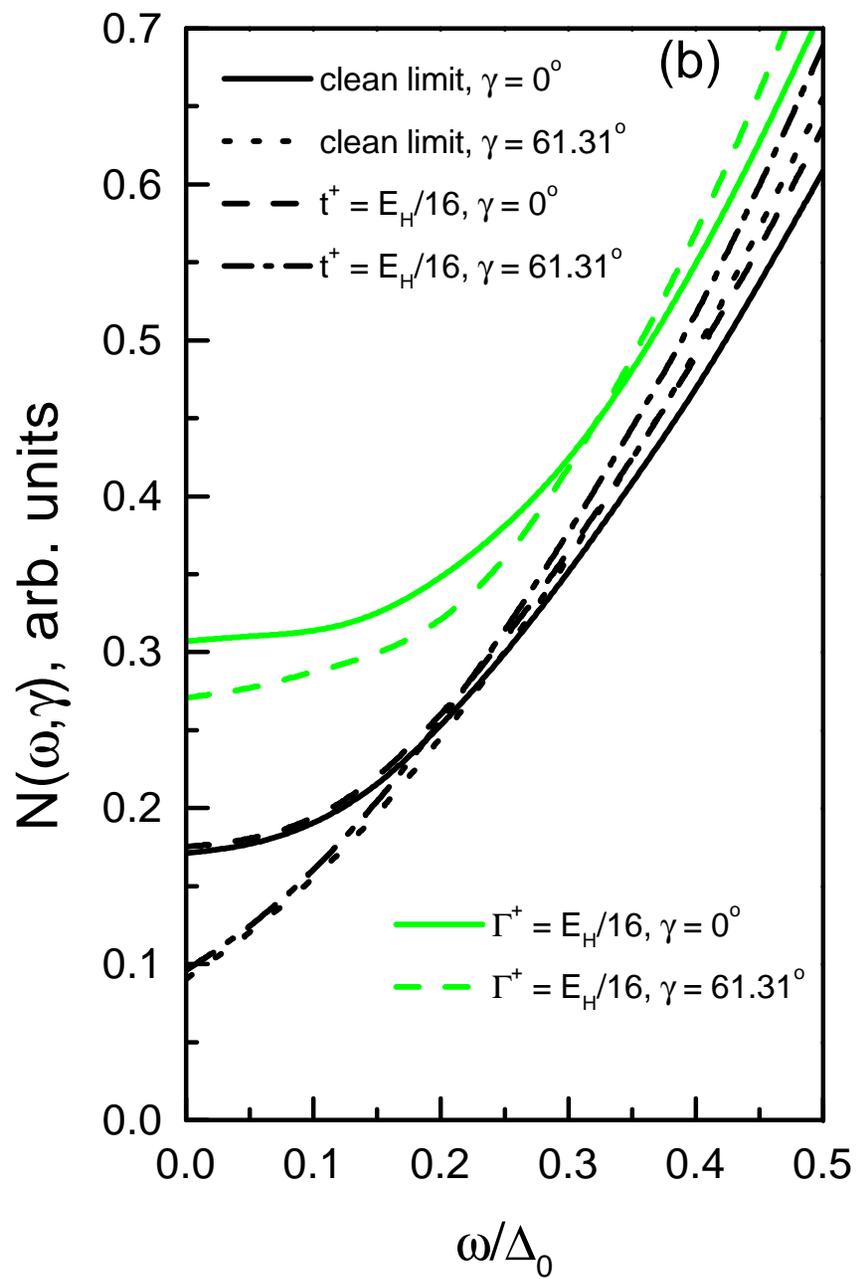

More insight into the frequency dependence of the anisotropy can be
obtained from the frequency dependence $N(\omega,\gamma)$ which is shown
for several fixed directions $\gamma$ in Fig.~\ref{f10}, frame (a)
and (b). These apply respectively to the tetragonal and orthorhombic
case. In frame (a) $\gamma = 0^\circ$ (antinodal direction) and
$\gamma = 45^\circ$ (nodal direction) are considered. For the pure
case the solid curve which applies for $\gamma = 0^\circ$, shows an
$\omega^2$ dependence at small $\omega$ while the dotted curve for
$\gamma = 45^\circ$ is linear as expected\cite{Vek1,Vek2} and falls
below the value along the antinodal direction. The nodal direction
is indeed the direction of the minimum in $N(0,\gamma)$ of Fig.~\ref{f6}.
As the frequency $\omega$ is increased the solid and dotted curves
rapidly come together and even cross so that the nodal direction
has a larger value of $N(\omega,\gamma)$ than the antinodal direction
for $\omega$ larger than the crossover frequency. The anisotropy is
now very small and will not be observable in specific heat experiments.
Dependencies of $N(\omega,\gamma)$ vs. $\omega$ are changed only
slightly with the introduction of Born scattering impurities; the
dotted curve is for $\gamma = 0^\circ$ and the dashed-dotted one for
$\gamma = 45^\circ$, and $t^+ = E_H/16$. On the other hand, the gray
curves for resonant scattering with $\Gamma^+ = E_H/16$ are very
different and are approximately constant at low $\omega$ 
with little difference between the
two directions. It is obvious that little anisotropy remains
in the QDOS. Similar results are given in frame (b) for
the orthorhombic case with $\alpha=0.41$, $s=-0.25$, and $\nu = 0.1$,
{\it i.e.} the same value of $E_H$ to gap amplitude is used as in
frame (a). The most important difference to be noted is that the
critical angle, at which an $\omega$ dependence is found in the clean
limit, is at the angle $\gamma = 61.3^\circ$ and not at $\gamma = 45^\circ$.
Also, the anisotropy is not washed out quite as much in the resonant
scattering case when compared to the tetragonal case. Further in this
case at sufficiently high frequencies the two curves for $\gamma = 0^\circ$
and $\gamma = 61.3^\circ$ cross and the curve for $\gamma = 0^\circ$
falls below the one for $\gamma = 61.3^\circ$ thus reversing the pattern
of maxima and minima shown in Fig.~\ref{f9} for lower values of
$\omega$. We also note that as $\omega$ increases out of zero the QDOS
$N(\omega,\gamma)$ decreases for both $\gamma$ values given in frame (a)
(tetragonal case) while for the orthorhombic case in frame (b) the reverse
is true. This effect is also seen in Figs.~\ref{f7} and \ref{f9}. In
the tetragonal case the solid line of Fig.~\ref{f7} for which
$\omega = 0$ is above the others for $\omega\ne 0$ while in Fig.~\ref{f9},
the orthorhombic case, it is below.

\section{Conclusions}

We have calculated the effect of the circulating vortex supercurrents on the
QDOS of a two dimensional $d$-wave superconductor. For the magnetic field
{\bf H} placed in the CuO$_2$ plane the QDOS varies as a function of the
angle $\gamma$ between {\bf H} and the ${\bf\hat{x}}$-axis. We have
included in our calculations an $a$-$b$ effective mass anisotropy of
about 2 with a view of modeling the band structure of YBCO which is 
orthorhombic because of
the existence of CuO chains, and for which the charge carrier oscillator
strength on the chains is approximately of the same order as in the
CuO$_2$ planes. In addition, the gap on the Fermi surface was assumed
to posses a subdominant $s$-wave as well as a dominant $d$-wave character.
Consideration of penetration depth data fixes the value of this admixture so that
we have no adjustable parameters. Impurities are also included in our
calculations in both, Born's and unitary limits. It is found
that the four-fold pattern predicted for the dependence of the zero
frequency value of the QDOS in a pure tetragonal system is now changed
to a two-fold variation due to the orthorhombicity in the gap or in the
band structure. The introduction of impurities leaves the symmetry of
the pattern unchanged but can reduce its amplitude. It is also found that
the pattern of variation of $N(\omega,\gamma)$ with $\gamma$ is altered
as the frequency is increased, although it remains four-fold for
tetragonal systems and two-fold for the orthorhombic case.

An important result is that the angle at which minima occur in $N(0,\gamma)$
vs. $\gamma$ does not correspond to the angle at which the zeros occur in
the gap as would be the case in a tetragonal system. Instead, the
minima occur when the normal state quasiparticle Fermi velocity in the
nodal direction is perpendicular to the direction of the vortex
supercurrents, {\it i.e.} parallel to the direction of the external
magnetic field.

Specific heat measurements in clean samples are most favorable for
observing the predicted two-fold pattern of anisotropy. To observe it
in YBCO detwinned samples are best because an average over equal numbers
of twins greatly reduces the predicted anisotropy for detwinned samples.
Consideration of more realistic tight binding band structure models
are in progress. Our very simple effective mass model, however, has
allowed us to understand the important changes orthorhombicity brings
to the magnetic field orientation dependence of the QDOS for a two
dimensional CuO$_2$ plane with {\bf H} in the plane. It has helped us
to determine the experimental conditions most favorable to the
observation of the predicted anisotropies.

\section*{Acknowledgments}

We have benefited from discussions with our colleagues P.J.\ Hirschfeld,
E.\ Nicol, I.\ Vekhter, and W.C.\ Wu.
Work supported in part by the Natural Sciences and Engineering Research
Council of Canada (NSERC), the Canadian Institute for Advanced
Research (CIAR), and by
Fonds zur F\"orderung der wissenschaftlichen Forschung (FWF), Vienna,
Austria under contract No. P11890-NAW.

\newpage
\begin{figure}
\caption{The ellipsoidal Fermi surface in {\bf k}-space given by
$\varepsilon_{\bf k} = 0$ in Eq.~(\ref{e1}) (solid line). Here we have
used $\alpha = 0.4$. Also shown is the vector {\bf k} with the
direction $\varphi$ and the corresponding Fermi velocity at {\bf k},
${\bf v}_F({\bf k})$. Note that it is not parallel to {\bf k}. The
transformed Fermi surface to {\bf p}-space is a circle and is shown
as a solid, gray line. $p_F = \sqrt{2\bar{m}\varepsilon_F}$, with
$\bar{m} = (m_x+m_y)/2$ and the angle $\phi$ is identified.}
\label{f1}
\end{figure}
\begin{figure}
\caption{Order parameters on the transformed Fermi surface ({\bf p}-frame)
as given by Eq.~(\ref{e7}). (a) $\alpha = s = 0$ (pure $d$-wave gap);
(b) $\alpha = 0.4$, $s=0$; (c) $\alpha = s = 0.4$; (d) $\alpha = 0.4$,
$s=-0.4$. The straight lines denote the nodal angles.}
\label{f2}
\end{figure}
\begin{figure}
\caption{The zero frequency value of QDOS, $N(0,\gamma)$, as a function
of magnetic field orientation in the CuO$_2$ plane with $\gamma$ the
angle of {\bf H} with respect to the $k_x$-axis. The solid line is
based on the approximate analytic formula (\ref{e24}) while the dashed
line is obtained from numerical evaluation of the exact one dimensional
integral (\ref{e18}). Only band anisotropy $(m_x\pm m_y)$ is included
with $\alpha = 0.2$ (ellipsoidal Fermi surface) and the gap is pure
$d$-wave with nodes in the laboratory, ({\bf k})-frame at $\pm\pi/4$
and symmetry related points. In the figure $\nu = E_H/\Delta_0 = 0.3$.}
\label{f3}
\end{figure}
\begin{figure}
\caption{The Fermi surface (ellipse) in the laboratory frame. Also
shown is the direction of the node ${\bf k}_n$ of the gap
$(\varphi_n)$. The direction of the magnetic field {\bf H} with the
angle $\gamma$ is taken to be parallel to the Fermi velocity at
the node ${\bf v}_F({\bf k}_n)$ so that the supercurrents defining the
vortex flow in a plane (indicated by the thick, solid, gray line) are
perpendicular to the $k_x$-$k_y$ plane and parallel to the tangent on
the Fermi surface at ${\bf v}_F({\bf k}_n)$.}
\label{f4}
\end{figure}
\begin{figure}
\caption{The QDOS $N(0,\gamma)$ at zero frequency as a function of
magnetic field orientation in the CuO$_2$ plane $(\gamma)$ for various
values of the band structure anisotropy $(\alpha)$. As $\alpha$
increases the amplitude of the oscillations increases: the solid line
is for $\alpha = 0.2$, the dashed one for $\alpha = 0.4$, the dotted
one for $\alpha = 0.6$, and, finally, the dashed-dotted one for
$\alpha = 0.8$.}
\label{f4New}
\end{figure}
\begin{figure}
\caption{The frequency dependence $(\omega)$ in units of $\Delta_0$
$(\omega/\Delta_0)$ of the QDOS, $N(\omega,\gamma)$ for the pure case
at four different magnetic field orientations, namely $\gamma = 0^\circ$
(solid line), $45^\circ$ (dotted line), $51^\circ$ (dashed line), and
$56^\circ$ (dashed-dotted). The $\gamma = 0^\circ$ curve shows $\omega^2$
behavior at low $\omega$ while the $\gamma=56^\circ$ is linear in
$\omega$ as would be the case in zero field for a $d$-wave
superconductor.}
\label{f5}
\end{figure}
\begin{figure}
\caption{The zero frequency value of the QDOS $N(0,\gamma)$ as a function
 of the
magnetic field orientation $\gamma$ for a pure $d$-wave tetragonal
$(\alpha = s = 0)$ superconductor. The solid line includes some
impurities characterized by $t^+ = E_H/16$ and the dashed curve is the
clean limit result.}
\label{f6}
\end{figure}
\begin{figure}
\caption{The dependence of the QDOS, $N(\omega,\gamma)$ on the orientation
of the magnetic field in the CuO$_2$ plane $(\gamma)$ for four different
frequencies $\omega$ in the unitary limit with $\Gamma^+ = E_H/16$. The
solid curve is for $\omega = 0$, the dashed one for $\omega/\Delta_0 = 0.05$,
dotted for $\omega/\Delta_0 = 0.1$ and dashed-dotted for
$\omega/\Delta_0 = 0.15$. $\nu = E_H/\Delta_0 = 0.1$ in a tetragonal
system with a pure $d$-wave gap.}
\label{f7}
\end{figure}
\begin{figure}
\caption{The dependence of the zero frequency QDOS, $N(0,\gamma)$, on
the magnetic field orientation $\gamma$ in the CuO$_2$ plane for an
orthorhombic system with $\alpha = 0.4$, $s = -0.25$, and 
$\nu = E_H/\Delta_0 = 0.1$. The dotted curve is for the pure system,
the solid one for Born scattering with $t^+=E_H/16$, and the solid gray
curve is for unitary scattering with $\Gamma^+=E_H/16$. The dashed-dotted
curve was obtained from the dotted one by taking the average of it, as is,
and its displacement by $90^\circ$; it applies to experiments that
average over equal numbers of twins. The dashed curve includes Born
scattering and the dashed gray curve unitary scattering. Both are for
twinned crystals.}
\label{f8}
\end{figure}
\begin{figure}
\caption{The dependence of the QDOS, $N(\omega,\gamma)$, on the
orientation of the magnetic field $\gamma$ in the CuO$_2$ plane for
different frequencies, namely $\omega/Delta_0 = 0$ (solid line),
$\omega/\Delta_0 = 0.05$ (dashed-dotted line), $\omega/\Delta_0 = 0.1$
(dashed line), $\omega/\Delta_0 = 0.15$ (dashed-double
dotted line), and $\omega/\Delta_0 = 0.2$ (dotted line). Orthorhombic
$d$-wave with $\alpha = 0.4$, $s=-0.25$, $\nu = 0.1$, and $\Gamma^+=%
E_H/16$.}
\label{f9}
\end{figure}
\end{document}